# Title: Giant magnetic anisotropy in the atomically thin van der Waals antiferromagnet FePS$_3$

Youjin Lee[1,2,#], Suhan Son[1,2,#], Chaebin Kim[1,2,#], Soonmin Kang[2], Junying Shen[3], Michel Kenzelmann[3], Bernard Delley[4], Tatiana Savchenko[5], Sergii Parchenko[5], Woongki Na[6], Ki-Young Choi[2], Wondong Kim[7], Hyeonsik Cheong[6], Peter M. Derlet[4], Armin Kleibert[5,*], and Je-Geun Park[1,2,*]

[1]Center for Quantum Materials, Seoul National University, Seoul 08826, Republic of Korea
[2]Department of Physics and Astronomy & Institute of Applied Physics, Seoul National University, Seoul 08826, Republic of Korea
[3]Laboratory for Neutron Scattering and Imaging, Paul Scherrer Institut, CH-5232 Villigen PSI, Switzerland
[4]Condensed Matter Theory Group, Paul Scherrer Institut, CH-5232 Villigen PSI, Switzerland
[5]Swiss Light Source, Paul Scherrer Institut, CH-5232 Villigen PSI, Switzerland
[6]Department of Physics, Sogang University, Seoul 04107, Republic of Korea
[7]Korea Research Institute of Standards and Science (KRISS), Daejeon 34113, Republic of Korea

\# Authors with equal contribution

$ Corresponding authors: armin.kleibert@psi.ch and jgpark10@snu.ac.kr

**Abstract**

Van der Waals (vdW) magnets are an ideal platform for tailoring two-dimensional (2D) magnetism with immense potential for spintronics applications and are intensively investigated. However, little is known about the microscopic origin of magnetic order in these antiferromagnetic systems. We used X-ray photoemission electron microscopy to address the electronic and magnetic properties of the vdW antiferromagnet FePS$_3$ down to the monolayer. Our experiments reveal a giant out-of-plane magnetic anisotropy of 22 meV per Fe ion, accompanied by unquenched magnetic orbital moments. Moreover, our calculations suggest that the Ising magnetism in FePS$_3$ is a visible manifestation of spin-orbit entanglement of the Fe 3$d$ electron system.

**Introduction**

The history of 2D magnetism is rich with exciting theoretical and experimental breakthroughs, starting from the Onsager solution of the 2D Ising Hamiltonian[1], which has given a strong theoretical basis and motivation to realise 2D magnetic systems. The discovery of magnetic van der Waals (vdW) materials is thus an important new development[2-3], which is also expected to play an essential role in much wider research fields such as nanotechnology and spintronics[4-5]. For instance, vdW magnetic materials offer a natural platform to investigate the fundamentals of 2D magnetism and their possible applications, including spintronics, by providing natural magnetic materials with atomic thickness. Specifically, the new class of vdW magnets provides the long sought-after experimental testbeds of the fundamental Hamiltonians of 2D magnetism: Ising, XY, and Heisenberg models[6]. The first experimental test of the Onsager solution of the Ising model was performed on antiferromagnetic $FePS_3$[7], followed by the discovery of 2D ferromagnetic van der Waals materials such as $CrGeTe_3$[8] and $CrI_3$[9]. Eventually, the XY and Heisenberg models were subsequently realized using $NiPS_3$[10-11], $CrCl_3$[12], and $MnPS_3$[13-15], respectively.

The investigation of vdW materials has so far focused on 2D magnetic order phenomena, and much less attention has been paid to their microscopic origin. Typically, the role of orbital magnetism is often neglected since the magnetic orbital moment is supposedly quenched in most *3d* transition metal-based magnetic materials via so-called crystal field splittings due to the interaction with the surrounding atoms. However, systems with reduced symmetry can, in principle, host sizeable magnetic orbital moments approaching those of the respective free atoms[16-17]. In such a case, the spin-orbit interaction can give rise to new phenomena such as spin-orbit entanglement, a manifestation of direct quantum effects in the condensed matter state[18]. Moreover, $FePS_3$ is – as we will show – a material whose properties are dominated by strong electron correlations. In spin-orbit entangled systems, electron spins and orbital motions are locked in an entangled Hilbert space, leading to novel magnetic phenomena such as Kitaev physics[19]. Realizing and tuning such an entangled state would provide fundamentally new functionality to magnetic materials. Thus far, spin-orbit entanglement has been primarily considered in compounds containing heavy elements such as Ru, Ir, and *4f* lanthanides[20]. However, such efforts have been relatively less vigorously made for *3d* transition metals, let alone the important class of 2D vdW magnets. At this point, one may be better reminded that a standard model of magnetism would predict an almost negligible effect of the spin-orbit coupling on the magnetism of the *3d* transition metal element simply because it is too small: it is on the order of 15 ~ 20 meV for Fe metal.

In this work, we shed important light on the role of orbital magnetism and, in particular, spin-orbit entanglement in the prototypical 2D Ising-type antiferromagnet $FePS_3$, which persists down to the monolayer limit. In addition to the fundamental interest, understanding antiferromagnetic 2D vdW materials such as $FePS_3$ is important for developing bourgeoning antiferromagnetic spintronics[21-22]. Thus far, the investigation of antiferromagnetic 2D vdW materials has been plagued by the lack of adequate experimental tools, which can directly probe the magnetic moments of few-layer thin antiferromagnetic samples at the nanometer scale[23]. This unfortunate situation about antiferromagnetic vdW materials is in stark contrast with atomically thin ferromagnetic materials, directly measured via the magneto-optical Kerr effect (MOKE)[8-9] and, more recently, diamond nitrogen-vacancy (NV) magnetometry[24]. To date, Raman[11, 14] and second harmonic generation (SHG)[25] techniques have been used to study the 2D magnetism of atomically thin antiferromagnetic vdW materials. A recent transport measurement using the spin Hall magnetoresistance (SMR) succeeded in measuring the sublattice magnetization of exfoliated $FePS_3$[26], which is another direction for the studies of vdW antiferromagnets. Although they prove practical, there are also limitations, as they cannot provide direct information about the microscopic origin of the antiferromagnetic properties.

Here, we use X-ray photoemission electron microscopy (XPEEM) in combination with X-ray absorption (XA) spectroscopy and the X-ray magnetic linear dichroism (XMLD) effect at the Fe $L_3$ edge to investigate the electronic and magnetic state of $FePS_3$ with atomic monolayer sensitivity and nanometer spatial resolution[27-29]. $FePS_3$ exhibits antiferromagnetic Ising-type order down to the monolayer regime, as demonstrated using Raman spectroscopy[7]. Here, we show that the Ising-type behavior is due to a giant magnetic single-ion anisotropy, which stabilizes the out-of-plane orientation of the magnetic moments even far above the Néel temperature. As the origin of this extraordinarily high anisotropy, we identify large unquenched orbital moments of approximately 1 $\mu_B$ per Fe ion. Atomistic multiplet calculations reveal the exotic nature of the ground state of $Fe^{2+}$ ions in $FePS_3$. Despite the smallness of the spin-orbit coupling strength, we find that the ground state of the six $3d$ electrons in $FePS_3$ is characterized by completely mixed spin and orbital wave functions. Such an unfactorizable mixture of spin and orbital states is a manifestation of multi-electron spin-orbit entanglement. Our results demonstrate that quantum spin-orbit entanglement should be considered an essential part of 2D vdW magnets to explore their full potential for fundamental research and applications.

$FePS_3$ is an Ising-type antiferromagnet with spin moments parallel to the $c^*$-axis[30] in a honeycomb lattice with a zig-zag configuration (Fig. 1a). Magnetic susceptibility measurements and neutron scattering data revealed a large magnetic out-of-plane anisotropy in bulk $FePS_3$[30-31]. The Ising-type behavior and the magnetic easy axis were confirmed using a torque magnetometer for bulk $FePS_3$[32]. The strong out-of-plane anisotropy, in principle, should be sufficient to enable Ising-type magnetic order in the individual 2D $FePS_3$ layers[1]. Indeed, stable antiferromagnetic order was observed in a single monolayer of $FePS_3$ using Raman spectroscopy[7]. However, the degree of magnetic order and the role of the interface with the substrate remain unclear because of the rather indirect spectroscopic Raman detection of antiferromagnetic order via magnon excitation and the related complex interplay between photons, phonons, and magnons[33]. Moreover, the microscopic origin of the out-of-plane anisotropy remains unclear despite its fundamental importance. Interestingly, recent theoretical work suggested that the out-of-plane anisotropy might be associated with significant magnetic orbital moments in $FePS_3$[34]. Although the density functional (DFT) calculation found a sizeable magnetic anisotropy by including spin-orbit coupling, experimental verification remained elusive. More importantly, the possibly entangled nature of the electronic ground state is completely unexamined.

## XPEEM investigations of the Ising magnet FePS$_3$

To investigate the magnetic and electronic properties of FePS$_3$ with monolayer sensitivity we perform polarization-dependent XA spectroscopy and microscopy at the Fe $L_{2,3}$ edge employing XPEEM in the experimental geometry shown in Fig. 1b. In this geometry the difference in the XA spectra recorded with the two orthogonal polarizations ($\mathbf{E}_\parallel$ and $\mathbf{E}_\perp$, respectively), i.e. the XMLD effect, is a measure of the local antiferromagnetic order. The normalized XMLD asymmetry $X_M$ (see Fig. 2a) is proportional to $<M_z^2>-<M_x^2>$, with $<M_{z,x}^2>$ being the expectation values of the squared order parameter perpendicular or parallel to the sample plane (see Methods). Before the experiment, FePS$_3$ flakes are exfoliated onto an indium tin oxide (ITO) substrate, which serves as electrically conductive support for the XPEEM investigations. Fig. 1c shows an atomic force microscopy (AFM) image of a typical FePS$_3$ flake with micrometer-sized flat regions ranging from 1 to 20 monolayers (ML). The corresponding XPEEM elemental contrast map is displayed in Fig. 1d. The XA spectra in Fig. 1e were extracted from a 20 ML region of the FePS$_3$ sample for the two orthogonal polarizations recorded at 65 K, well below the Néel temperature $T_N$ = 118 K of bulk FePS$_3$. The polarization-dependent XA spectra exhibit three distinct features, denoted as A, B, and C, associated with the XMLD effect due to the magnetically ordered state below $T_N$. Finally, the local magnetic order of the FePS$_3$ flake is visualized using XMLD contrast maps (Fig. 1f). The figure shows the magnetic order to persist down to the monolayer (see also Fig. S5).

## Comparison with multiplet calculations and Monte Carlo simulations

We start our analysis by comparing the temperature-dependent XPEEM data of 20 ML FePS$_3$ with multiplet calculations and Monte Carlo simulations. Figs. 2a & 2b show the experimental XA spectra of 20 ML FePS$_3$ recorded at 65 and 300 K below and above $T_N$, respectively (see Methods for details). The linear dichroism in the spectra is significantly larger below $T_N$ than in the paramagnetic phase. The small remaining dichroic signal in the paramagnetic state reflects a non-magnetic contribution due to the low symmetry of the trigonally distorted FeS$_6$ octahedra in FePS$_3$, present in all temperature regions.

To obtain further insight into the origin of the magnetism and magnetic order of FePS$_3$, we simulated the XA spectra using ligand field multiplet theory (see Methods). As shown in Figs. 2a & 2b, our multiplet calculations performed for 65 and 300 K successfully reproduce the key features of the experimental XA and XMLD spectra. The multiplet calculations yield a spin moment $<S_z>$ = 1.75 $\pm$ 0.04 and a large orbital moment $<L_z>$ = 1.02 $\pm$ 0.04 per Fe ion, consistent with earlier DFT results[34]. Notably, these numbers compare well to those of Fe$^{2+}$ ions in a $d^6$ high spin state with unquenched $S$ = 2 and $\tilde{L}$ = 1. For such a case, significant magnetic orbital moments are expected to give rise to a high magnetic anisotropy[35].

To assess the magnetic anisotropy in FePS$_3$, we evaluate $X_M$ as a function of temperature and compare these experimental data to the $<M_z^2>-<M_x^2>$ obtained from ensemble-averaged Monte-Carlo (MC) simulations, as shown in Fig. 3a, for 20 ML. The MC simulations are based on a classical 2D honeycomb lattice of the Heisenberg exchange model with strong local out-of-plane anisotropy (see Methods). Using an intra-ML exchange coupling corresponding to a ground state out-of-plane exchange field of $J = 15$ meV and an out-of-plane single-ion anisotropy of $K = 22$ meV, we successfully reproduce the known Néel temperature of 115 K and the experimental XMLD amplitude data, $X_M$ (see Fig. 3a). The exchange and anisotropy values compare well to those deduced from earlier theoretical predictions[34, 36]. The high magnetic anisotropy is further consistent with the anisotropic magnetic susceptibility seen in bulk $FePS_3$ above $T_N$ (Fig. S1). The positive sign of $K$ further confirms the out-of-plane orientation of the magnetic moments, suggesting that the Ising-type magnetic order is an intrinsic property of the 2D layers of bulk $FePS_3$. It should be noted that the magnitude of the magnetic anisotropy estimated above is much larger, if not one of the largest, than in most other magnetic materials[16].

**Thickness dependence of magnetic anisotropy**

By taking advantage of the spatial resolution and sensitivity of XPEEM, we apply the same magnetic anisotropy analysis to sample regions with different thicknesses to reveal the role of interlayer and interface interactions. Fitting the temperature-dependent experimental data $X_M$ with calculated $<M_z^2> - <M_x^2>$ with varying $K$ at fixed $J$ (Fig. S3 and S4), we find practically the same bulk-like behavior and properties persisting down to 4 ML. Below 4 ML, we observe a significant drop in the normalized asymmetry $X_M$ at low temperature indicating some kind of change in the magnetic properties. The change in the magnetic properties are accompanied with changes in the XA spectra, which are most apparent in the monolayer (Fig 3c), which we can assign to an electron transfer from the ITO substrate to the first $FePS_3$ layers as discussed further below (see also the Supplementary Information). We also noticed some x-ray-induced modifications in the XA spectra for thicknesses below 3 ML, which prevents the temperature-dependent analysis of $X_M$. In these cases, we may extract the magnetic anisotropy parameter $K$ from Bruno's theory[35], which links the anisotropy energy to $\frac{1}{4}\xi\hat{S}\cdot(\Delta L)$,[37] where for the present case the orbital moment anisotropy $\Delta L = (L_z - L_x)$ and spin moment $\hat{S}$ are obtained from the multiplet calculations by fitting the respective XA spectra at low temperature. Here, $\xi$ is an atomic spin-orbit coupling constant, which has a value of 50 meV for Fe. Interestingly, we find that the $K$ values obtained from Bruno model are similar with what we got from MC down to 4 ML (see the grey diamonds in Fig 3b). For the 3ML, we find only a small decrease of $K$ in Bruno's model, 20.6±0.3 meV/Fe, as compared to the large drop of $K$, 11 meV/Fe from the MC calculations. For the monolayer, Bruno model yields $K \sim 11.6\pm0.4$ meV with $\Delta L$ and $\hat{S}$ from the data shown in Fig 3c. We argue that this reduction in $K$ in the lower thicknesses indicates an interface effect with the ITO, likely related with the electron transfer to the $FePS_3$.

The latter occurs due to a work function mismatch between the p-type semiconductor $FePS_3$[38] and the metallic ITO[39]. Including a charge transfer in our multiplet simulations is achieved by reducing the charge transfer energy $\Delta$, the effective coupling potential $V_{e_g^\sigma}$, $V_{a_{1g}}$, and $V_{e_g^\pi}$ in $H_{lmct}$ and ligand's crystal field $H_L$ parameters (other parameters are kept the same as that of 20 ML). Such simulations indeed reproduce the changes in the XA spectra for the monolayer (Fig.

3c) and further yield $<S_z>$ = 1.29± 0.01 and $<L_z>$=0.72±0.02, matching the reduced $X_M$ in the XMLD data. This reduction of the magnetic moments correlates well with the strongly reduced antiferromagnetic order peaks $P_1$ and $P_2$ in the Raman spectra of FePS$_3$ on the ITO substrate (Fig. 3d). At the same time, these data show that care must be exercised when bringing 2D vdW materials in contact with a substrate. Similar effects are well known from other systems, such as supported molecules, e.g.[40]. The present case demonstrates that electron transfer effect is also relevant in 2D vdW materials on supports, but might have been overlooked thus far, because the commonly used experimental techniques lack the required sensitivity to such effects. In the present case, electron transfer from ITO to FePS$_3$ modifies the electronic and magnetic properties of the first few layers.

**Spin-orbit entanglement**

We note here that upon introducing spin-orbit coupling, the $a_{1g}$ and $e_g^\pi$ states are spin-orbit entangled (Fig. 4b). Indeed, our multielectron calculations of FePS$_3$ reveal that the spin-orbit interaction leads to the spin-orbit entangled ground state of the $3d$ electronic system. In the multielectron calculations, we took a total Hamiltonian for the $d^6$ electrons of Fe$^{2+}$ for the trigonal symmetry as detailed in the SI. We start the multielectron calculation from the cubic crystal electric field $H_{CEF}$ with basis of $|L = 2, m_L\rangle$, because crystal electric field acts only on orbital moment. In our diagonalization of cubic $H_{CEF}$, we can start the calculation with the ground orbital triplet $^5T_{2g}$ state (i.e., $\tilde{L}$ = 1 and $S$ = 2) because the orbital moment of excited state $E_g$ is zero and the energy splitting from $T_{2g}$ to $E_g$ is an order of eV. With the ground state $^5T_{2g}$, the simplified Hamiltonian can be described as $H = \lambda \tilde{L} \cdot S + \Delta_{trig}(\tilde{L}_z - \frac{2}{3})$, where $\lambda$ is the spin-orbit coupling and $\Delta_{trig}$ is the trigonal distortion. Using the 15 basis functions of $(2\tilde{L} + 1) \times (2S + 1)$ in the $|\tilde{L}, S\rangle = |\tilde{L} = 1\rangle_L \otimes |S = 2\rangle_S$ basis, we calculate the ground state of FePS$_3$ by numerically diagonalizing the 15×15 matrix. By setting the $\lambda$ = 13 meV and $\Delta_{trig}$ = -10 meV from Quanty's best fit parameter value, multielectron calculation then produces a ground state of Fe$^{2+}$ with the wave function of $|\Psi\rangle = \pm 0.8325|\mp 1\rangle_L \otimes |\pm 2\rangle_S \mp 0.4712|0\rangle_L \otimes |\pm 1\rangle_S \pm 0.2914|\pm 1\rangle_L \otimes |0\rangle_S$, given in the $|L,S\rangle$ basis functions of $\tilde{L}$ = 1 and $S$ = 2 for the six $3d$ electrons of Fe$^{2+}$. This ground state wave function cannot be factorized due to spin-orbit entanglement. Such complete mixing of all $L$-$S$ basis functions is definitive evidence of spin-orbit entanglement for the ground state wave function. The entanglement would naturally couple the spin moment and the orbital moment, reinforcing the large magnetic anisotropy. We note that this wavefunction also has a large orbital moment of $<L_z>$=0.6081, which is close to the values from our multiplet calculations (see discussions further above) and consistent with the recent LDA+U calculations[34]. To confirm our conclusion about the spin-orbit entangled ground state, we analyzed the branching ratio of $I_{L3}/I_{L2}$ of the experimental XA spectrum at 65 K. The quantity $I(L_{2,3})$ is the integrated intensity of the "white line," (colored area in Fig. S8) an intense absorption spectrum in the Fe $L_{2,3}$ edge. Traditionally, the branching ratio of XAS data has been used as a strict test of spin-orbit entanglement[41-42]. As detailed in the SI, the branching ratio of our experimental data is 3.84, significantly higher than a case without entanglement, $I_{L3}/I_{L2}$ = 2. We think this constitutes compelling evidence supporting the spin-orbit entanglement in FePS$_3$. Interestingly, the wavefunctions of the ground state have a distinct shape, as shown in Fig. 4b, with $e_g^\pi$ having more prominent in-plane orbital moments.

At first glance, a spin-orbit entangled ground state in FePS$_3$ appears to be a surprising result, as the spin-orbit coupling is much weaker for *3d* transition metal elements than systems based on Ir, Ru, and *4f* elements. However, it is known that for certain conditions, even weak spin-orbit coupling can easily give rise to ground states with significant spin-orbit entanglement for several *3d* transition metal elements. For instance, $Co^{2+}$ ions are a case in point, as discussed in Ref.[43]. Another example is an Ising metallic system $Fe_{1/4}TaS_2$[17]. This system also has a large magnetic anisotropy with unquenched orbital moments. The large magnetic anisotropy ~14 meV/Fe originates from the same mechanism as FePS$_3$, combination of spin-orbit coupling and unquenched orbital moment ~ 1 $\mu_B$/Fe from trigonal distortion. Finally, additional optical data exhibit visible temperature-dependent spectral transfer in FePS$_3$ (see Fig. S9). A similar spectral transfer was recently considered evidence of a strong correlation in NiPS$_3$[44]. Thus, finding a spin-orbit entangled state in *3d* transition metal compounds such as FePS$_3$ provides prominent new opportunities for investigating the potential effect of strong electron correlation on the final entangled ground state, most notably all on the 2D limit. It is entirely an open question how strong electron correlation impacts the spin-orbit entangled ground state of FePS$_3$, which will be the subject of future studies.

**Conclusion**

Our experiments demonstrate that orbital magnetism and spin-orbit entanglement play a key role in the enormous magnetic anisotropy and Ising-type magnetism of FePS$_3$. We obtain the sizeable magnetic anisotropy value of 22 meV/Fe from our analysis and demonstrate the existence of an unquenched orbital moment from multiplet calculations by fitting polarization- and temperature-dependent XA data. With the high-resolution XMLD mapping results, we can analyze the magnetic anisotropy's thickness dependence down to monolayer FePS$_3$. It is also crucial to note that the gigantic magnetic anisotropy is due to spin-orbit entanglement for $Fe^{2+}$ under trigonal elongation: the Ising magnetism of FePS$_3$ is intrinsically a quantum phase with strong entanglement. Finally, we demonstrate an electron transfer effect at the interface between the ITO substrate and FePS$_3$. Being able to probe such effects is crucial for potential applications of thin 2D vdW material layers. Therefore, we anticipate that our results will open the door to new investigations into quantum phases in 2D vdW magnets.

**Methods**

Sample preparation and basic characterization

FePS$_3$ single crystals are synthesized employing chemical vapor transport (CVT). The Fe, P, and S powders (99.99%, Alfa Aesar) were sealed into a quartz ampule. The sample is then placed in a horizontal two-zone furnace with temperatures of 750 °C (hot zone) and 730 °C (cold zone) and kept for 9 days. The magnetic susceptibility of bulk FePS$_3$ was measured using a commercial magnetometer (MPMS5, Quantum Design) (Fig. S1). The measurement is performed by field cooling with a 300 Oe magnetic field. Au markers are lithographically patterned on the ITO substrates and used to identify and locate selected FePS$_3$ flakes for correlating the same sample's XPEEM and atomic force microscopy (AFM). FePS$_3$ flakes were mechanically exfoliated on an ITO (70 nm)/Si substrate, and AFM was used to determine the thickness of the exfoliated flakes. Exfoliation and AFM measurements were carried out in a glove box filled with Ar gas.

Before the main experiment of FePS$_3$, we did the feasibility test of XMLD-PEEM on different substrates. Since the FePS$_3$ is insulating, eliminating the charging effect is crucial for obtaining the correct x-ray absorption coefficient. We tried the ITO substrate and SiO$_2$ substrate with a patterned gold window. In test experiments, the ITO substrate gave more stable scanning images than the gold pattered SiO$_2$ substrate.

Raman measurement

Raman scattering measurements are carried out on the sample fabricated by the same method. We employ an Ar-ion laser with a wavelength of 488 nm (2.54 eV) and a power of ~50 μW. The substrate with the exfoliated samples is loaded into a He-flow optical cryostat (Oxford MicrostatHe2). The scattered light from the sample was dispersed using a Jobin-Yvon Horiba iHR550 spectrometer (2400 grooves/mm) and detected with a liquid nitrogen-cooled CCD.

X-ray absorption and magnetic linear dichroism spectroscopy

The X-PEEM experiments are performed at the Surface/Interface Microscopy (SIM) beamline of the Swiss Light Source, Paul Scherrer Institute. Element-specific magnetic information at a spatial resolution of 50 – 100 nm is obtained by tuning the X-ray photon energy to the *L$_3$* absorption edge of Fe. The experiment is performed from low to high temperatures to minimize X-ray-induced sample damage. ImageJ is used for image processing steps such as drift correction and normalization to extract the intensity from the region of interest and pixel-wise contrast calculations.

Recording sequences of XPEEM images at successive photon energies and as a function of polarization enables us to acquire thickness- and polarization-dependent XA spectra to probe the anisotropic properties of the sample. The XA spectra are acquired by sequencing XA intensities from the XPEEM images at successive photon energies. A linear baseline correction was performed for each spectrum. Normalization is achieved by dividing each data point by the $L_3$ peak intensity. XPEEM elemental contrast maps are obtained by pixelwise division of images recorded at the Fe $L_3$ peak ($h\nu$ = 708.5 eV) and pre-edge energy ($h\nu$ = 704.2 eV)

Magnetic contrast maps are obtained by pixel-wise evaluation of the normalized XMLD asymmetry $A = \left[I_\perp^{(C,B)} - I_\parallel^{(C,B)}\right] / \left[I_\perp^{(C,B)} + I_\parallel^{(C,B)}\right]$, where $I_\perp^{(C,B)}$ and $I_\parallel^{(C,B)}$ denote the ratio of the local absorption intensities at the respective photon energy for two different polarizations. The asymmetry is proportional to $<M_z^2> - <M_x^2>$, with $<M_{z,x}^2>$ being the expectation values of the squared order parameter with components perpendicular or parallel to the sample plane. Hence, a disordered state (static or dynamic) with $<M_{z,x}^2> = 0$ yields $A = 0$, while $A \neq 0$ indicates a common orientation of the atomic magnetic moments.

The magnetization- and polarization-dependent XA intensity can be described as $I(\mathbf{M},E) = I_0 + c\,(\mathbf{E}\cdot\mathbf{M})^2$, with $I_0\,(h\nu)$ being the isotropic, photon energy $h\nu$ dependent XA, $\mathbf{E}$ being the polarization vector of linear polarized X-rays, $\mathbf{M}$ being the local order parameter, and $c = c\,(h\nu)$ being a photon energy-dependent parameter containing information about the microscopic magnetic properties of the sample[45]. The order parameter of FePS$_3$ is given by $\mathbf{M} = (\mathbf{M}_\uparrow - \mathbf{M}_\downarrow)/2$, where $\mathbf{M}_{\uparrow\downarrow}$ corresponds to the total probed magnetization of the oppositely polarized Fe ion sublattices in the zig-zag chains. XPEEM images were taken with the X-ray polarization either parallel, $\mathbf{E}_\parallel = (E,0,0)$, or almost perpendicular to the sample plane, $\mathbf{E}_\perp = (0, E\cdot\sin\theta, E\cdot\cos\theta)$, and with $\theta = 16°$ being the angle of incidence of the X-rays (Fig. 1b).

Simulation of XA spectra

The quantum many-body program Quanty[46] is used to simulate XA spectra. The script language Quanty enables us to calculate x-ray absorption spectra, defining operators in second quantization and calculating the eigenstates of Green's functions for these operators[47]. From calculated the eigenstates, we can get the expectation value of ground state including $<L_{x,y,z}>$ and $<S_{x,y,z}>$. Quanty input files are generated by Crispy[48], a graphical user interface core-level spectra simulation program. A ligand field multiplet model is employed, considering the *pd*-hybridization of Fe *d*-orbitals and S *p*-orbitals in the FeS$_6$ cluster[46]. To reflect the trigonal elongation along the [111] direction of the FeS$_6$ cluster, we add $D_{3d}$ Hamiltonian terms to the original Crispy input file. The calculated geometry, including the incident X-ray direction and E-field vector orientation, is shown in Fig. 4a. The input parameters are free parameters, so we fit their values to reproduce the experimental XA spectrum. The crystal field $H_c$ and the *pd* hybridization terms, $H_{lmct}$ and $H_L$, are adjusted to match the experimental XA spectra in the paramagnetic state at 300 K. The trigonal distortion parameter $D_{t2g}$ is assigned to be -10 meV from the previous optical result[49]. An exchange field is applied parallel to the $c^*$-axis direction to reproduce the magnetic state of FePS$_3$ below $T_N$. Other parameters are fixed to the parameters fitted at 300 K. This exchange field acts only on the spin moment and is defined as $H_{ex} = \sum_i J\langle S_i\rangle$[50]. Increasing the exchange field $H_{ex}$ along z-direction induces an increase in $<S_z>$ and $<L_z>$ due to spin-orbit coupling. The increased z-component of the spin and orbital moments leads to boosted magnetic anisotropy (Fig S2 a,b). In addition, the normalized asymmetry $X_M$ increases with a larger H$_{ex}$ (Fig. S2 c). The input parameters from the best fit are summarized in Table 1.

**Table 1**. Summary of parameters used for the Quanty simulation

| Hamiltonian | Parameters |
|---|---|
| Atomic terms<br>H$_d$ | U$_{dd}$ = 2.0 eV<br>U$_{pd}$ = 2.5 eV |
| Crystal field<br>H$_c$ | 10Dq = 0.4 eV<br>D$_{t2g}$ = -0.01 eV |
| Ligand-metal charge transfer<br>H$_{lmct}$ | Δ = 1.0 eV<br>V$_{a1g}$ = 2.3 eV<br>V$_{eg\pi}$ = 2.2 eV<br>V$_{eg\sigma}$ = 3.0 eV |
| Ligand crystal field<br>H$_L$ | 10Dq$^{lig}$ = 1.5 eV<br>D$_{t2g}^{lig}$ = -0.1 eV |

The Hamiltonian used in Quanty is $H_{tot} = H_d + H_c + H_{lmct} + H_L + H_{ex}$:

- Atomic terms H$_d$ are for the Fe electrons' Coulomb interactions and spin-orbit coupling.

  The on-site Coulomb repulsion in different *d-d* orbitals U$_{dd}$

  Coulomb interaction between the 2*p*-core hole and 3*d*-electron U$_{pd}$

  Slater integrals are reduced to 80% of the Hartree-Fock value with a spin-orbit coupling constant of $\xi = 0.05\ eV$.

- Crystal electric field H$_c$ reflects the splitting of Fe 3*d* orbitals by the crystal electric field (see Fig. 4b)

  Cubic crystal field 10Dq split Fe 3*d* orbitals into $t_{2g}$ and $e_g$ states.

  D$_{t2g}$ is a trigonal distortion parameter that splits the $t_{2g}$ orbitals into the $e_g^\pi$ and $a_{1g}$ states. D$_{t2g}$ is defined as E($e_g^\pi$) - E($a_{1g}$). The negative D$_{t2g}$ reflects the trigonal elongation, setting the doublet $e_g^\pi$ as a ground state.

  By 10Dq and D$_{t2g}$, the energies of the Fe 3*d* orbitals are described as

$$E_{e_g^\sigma} = 0.6 \times 10Dq \quad (Eq.1)$$

$$E_{a_{1g}} = -0.4 \times 10Dq - 2/3 \times D_{t2g} \quad (Eq.2)$$

$$E_{e_g^\pi} = -0.4 \times 10Dq + 2/3 \times D_{t2g} \quad (Eq.3)$$

- Ligand-metal charge transfer $H_{lmct}$ is the Hamiltonian that considers the charge transfer effect from the S ligand to the Fe 3$d$-metal.

> Charge transfer energy $\Delta$ is the energy difference between $d^6$ and $d^7\underline{L}$, where $\underline{L}$ denotes the S $p$-orbital hole states; $\Delta = E(d^7\underline{L}) - E(d^6)$.
>
> The hopping between the S and Fe ions can be depicted as an effective potential coupling of two different orbitals. This effective potential coupling describes the Fe 3$d$-orbital energy states of D$_{3d}$ symmetry: i.e. $V_{e_g}^\sigma$, $V_{a_{1g}}$, and $V_{e_g}^\pi$.

- $H_L$ is the sulfur ligand crystal electric field Hamiltonian.

> Cubic crystal field parameter 10Dq$^{lig}$ and trigonal distortion parameter $D_{t2g}^{lig}$

Monte-Carlo simulation

We investigated the magnetic properties through a Monte Carlo simulation. A classical 2D honeycomb lattice Heisenberg exchange model was employed with strong local out-of-plane anisotropy based on the quantum spin Hamiltonian of Refs.[30, 51]. The Heisenberg exchange coupling includes up to the 4$^{th}$ nearest neighbor. Following Ref.[51], we consider only intralayer exchange interactions: including an interlayer exchange interaction as in Ref.[30] results in a minor change. Thus, our classical vdW material is modeled as a stack of noninteracting MLs, each described by

$$H = -\frac{1}{2}\sum_{ij} J_{ij}\hat{m}_i \cdot \hat{m}_j - K\sum_i (\hat{m}_{z,i})^2 \quad (Eq.4)$$

A Néel temperature of 125 K is obtained with the exchange parameters $J_1$ = *14.6*, $J_2$ = *-0.4*, $J_3$ = *-9.6* and $J_4$ = *-0.073* meV and an out-of-plane anisotropy $K$ = *26.6* meV. The corresponding ground state exchange field is $H_{ex} \approx J_1 - 2J_2 - 3J_3 = 15\ meV$. We chose that the ratios of these parameters are equal to those of the quantum spin Hamiltonian found in Ref.[30] ($J^{quantum}M^2 \sim J$), resulting in a classical moment magnitude estimate $M$ compatible with that of Fe. For the study of the phase transition, a classical spin Hamiltonian is justified.

**Acknowledgments**

We thank Seokhwan Yun and Daniel Khomskii for the helpful discussions. Part of this work was performed at the Surface/Interface: Microscopy (SIM) beamline of the Swiss Light Source (SLS), Paul Scherrer Institut, Villigen, Switzerland.

**Funding:**

Leading Researcher Program of the National Research Foundation of Korea, Grant No. 2020R1A3B2079375 (YL, SS, SK, CK, J-GP)

The core center program of the Ministry of Education, Korea, Grant No. 2021R1A6C101B418 (J-GP)

Swiss National Foundation, Grant. No 200021_160186 (JS, MK, BD, TS, SP, PD, AK)

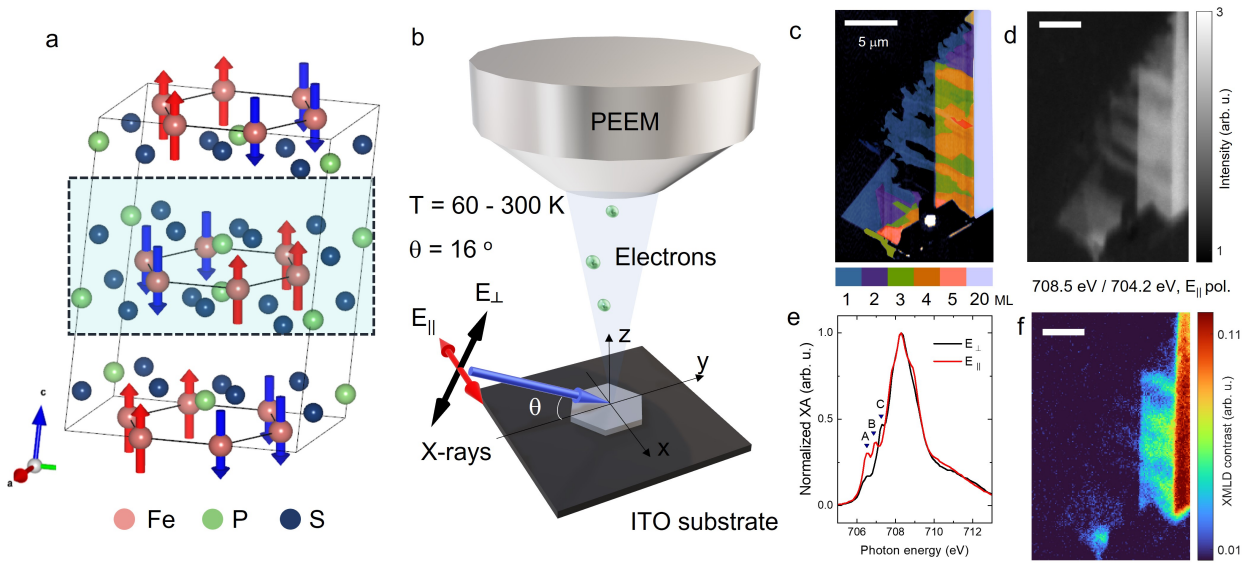

Fig. 1. **Schematic of the atomic structure of FePS$_3$ and the experimental setup. a**. The magnetic moments are ordered in a zig-zag chain direction parallel to the *c\**-axis (red and blue arrows). The jade-green shadow highlights the monolayer structure of FePS$_3$. **b**. Schematic of the XPEEM measurements with linearly polarized X-rays (E$_\parallel$ and E$\perp$) exciting the sample at a grazing angle (16°). **c**. False-color AFM image of the exfoliated FePS$_3$ flake on an ITO support. Each color corresponds to a given number of FePS$_3$ monolayers, as shown in the color bar. **d**. XPEEM elemental contrast map of the same FePS$_3$ flake. **e.** Normalized XA spectra at the Fe $L_3$ edge were extracted from a 20 ML thick region of the FePS$_3$ sample for both polarizations. Three pre-peaks are denoted as A, B, and C. **f**. Corresponding XMLD asymmetry map. The data in **d, e**, and **f** are obtained at 65 K, below T$_N$. The scale bars are 5 μm.

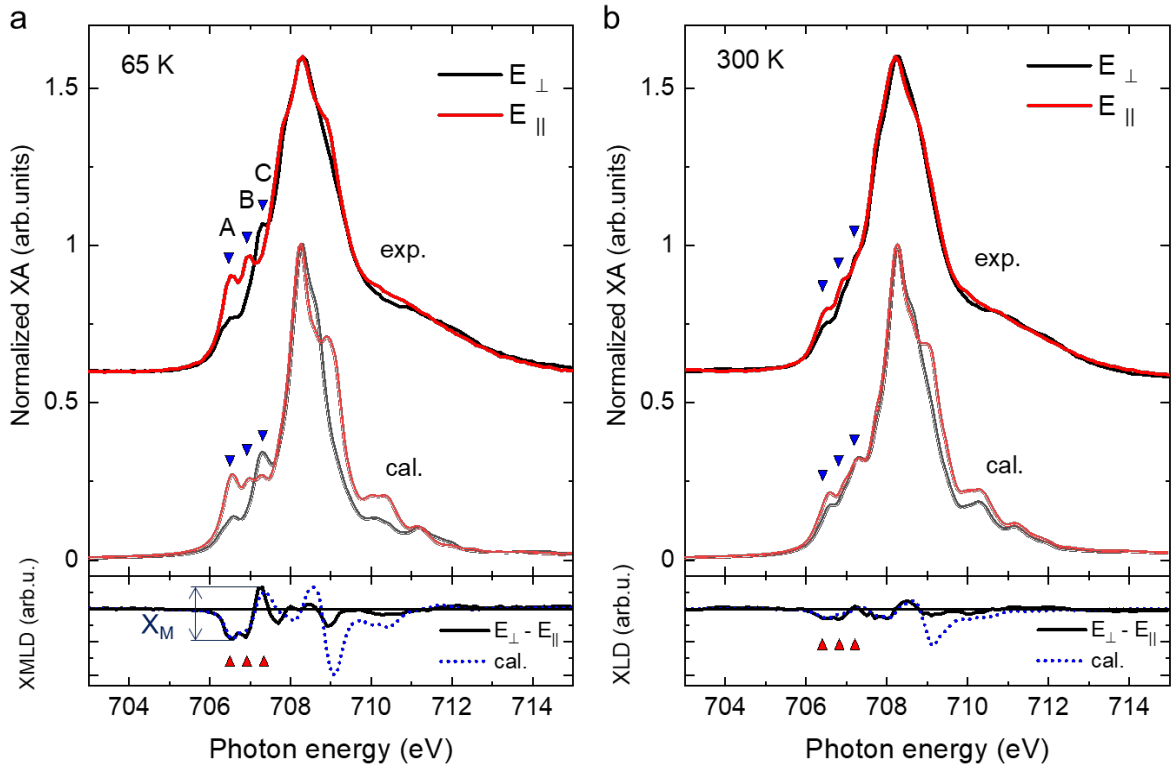

Fig. 2. **XA spectra of 20 ML-thick FePS$_3$.** Data are taken at **a**. 65 K and **b**. 300 K. The experimental (upper) and theoretical (lower) XA spectra for E$_\parallel$ (red) and E$_\perp$ (black). The experimental XA spectra are shifted by 0.6 upwards for a better comparison. The corresponding experimental (black) and calculated (blue dotted) XMLD spectra are shown at the bottom.

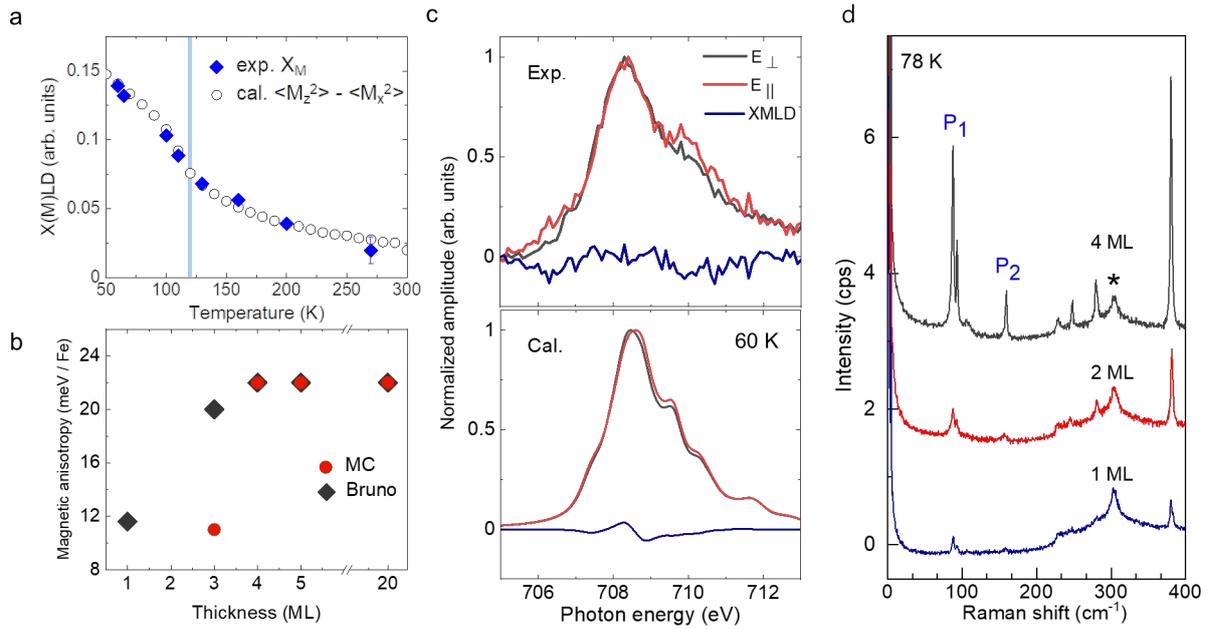

Fig. 3. **Thickness dependency of the magnetic anisotropy a.** Experimental XMLD amplitude $X_M$ (blue diamonds) together with $<M_z^2> - <M_x^2>$ from the Monte Carlo simulations (white circle) for 20 ML FePS$_3$ as a function of temperature. **b.** Magnetic anisotropy as a function of thickness extracted from Monte Carlo simulation (red circles) and Bruno's model (grey diamonds). **c.** Experimental (upper) and calculated (lower) monolayer XA and XMLD spectra. **d**. Raman spectra of FePS$_3$ flakes on the ITO/Si substrate. P1 and P2 originated from the long-range magnetic order, according to the previous Raman paper. The peak marked with an asterisk is a signal from the ITO substrate.

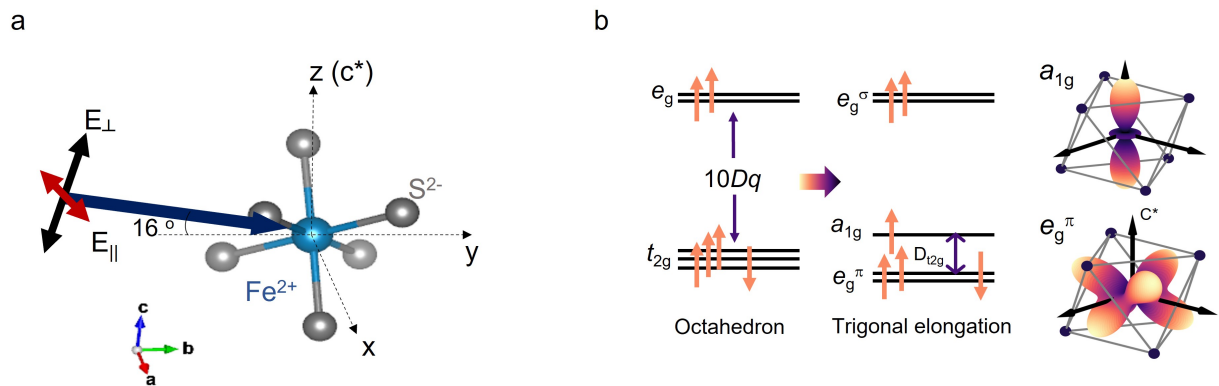

Fig. 4. **Simulated XA spectra with different spin orientations. a**. Schematic of the FeS$_6$ cluster and incident X-ray beam with linear polarization. The vectors of the incident X-ray beam and exchange field H$_{ex}$ in the multiplet calculation are determined from this scheme. **b.** The high spin configuration under the crystal field (left). The $e_g^\pi$ and $a_{1g}$ orbitals are visualized (right). Theoretical XA spectra with different spin orientations that are aligned

## Supplementary Information

**Magnetic anisotropy as a function of thickness**

A Monte-Carlo simulation protocol was performed to analyze the thickness dependence in FePS$_3$ few-layer for different out-of-plane anisotropy coefficients $K$, $K_N = K \times N$ with N ranging from 0.2 to 1. Thus, the quantity $<M_z^2> - <M_x^2>$ is obtained as a function of temperature for each $K_N$ (Fig. S3). Under the assumption of no inter-layer interaction, any thickness-dependent magnetic quantity must be related to the difference in the model system at the surface and/or the thin-flake/substrate interface compared to the bulk system. The present work considers the possibility of a modified out-of-plane anisotropy coefficient of the layer at the thin-flake/substrate interface. All the other layers of the flake are similar to that of the bulk. In such a scenario, the dependence of the *N* monolayer system can be derived from the weighted average of the *N-1* bulk layers and the single interface layer. Within this framework, the experimental data of Fig. 3a were fitted with a single free parameter being the appropriately reduced value of the interface layer, $K_N$.

We extracted the magnetic anisotropy value for thinner FePS$_3$ samples by the same method described above for the case of 20 ML. The magnetic anisotropy K was determined by fitting the experimental data $X_M$ with theoretical data $<M_z^2> - <M_x^2>$ calculated with different $K_N$ (Fig. S4). We could determine the magnetic anisotropy down to tri-layer; 4ML and 5ML have the same magnetic anisotropy of 22 meV/Fe with 20 ML, while it gets reduced to 11 meV/Fe for 3ML. In the case of bi- and monolayer, sample degradation by repetitive exposures hindered a complete analysis of temperature dependency.

**Multiplet calculation for the exact wavefunction of the electronic ground state in FePS$_3$.**

To better understand the entanglement nature of the ground state wavefunction in FePS$_3$, we use the crystal electric field (CEF) multiplet calculations [1, 2], which consider all 6 electrons of the Fe$^{2+}$ ion. Several works showed that the Fe$^{2+}$ free ion has a $^5D$ (i.e., $S = 2$, $L = 2$) multiplet ground state [1, 3-5]. We first consider the cubic CEF effect, the crystal electric field of octahedron symmetry, usually denoted as $10Dq$. We consider the $L = 2$ orbital moment in $^5D$ symmetry because the CEF effect impacts only the orbital moment. The Hamiltonian of a cubic crystal field $\hat{\mathcal{H}}_{CEF}$ can be written in terms of the Stevens operators $\hat{O}_4^0, \hat{O}_4^4$ and the numerical coefficient $B_4$ as

$$\hat{\mathcal{H}}_{CEF} = B_4 \left( \hat{O}_4^0 + 5\hat{O}_4^4 \right) \qquad (Eq.S1)$$

The numerical coefficient $B_4$ is defined as $\beta \langle r^4 \rangle$, where $\beta$ is the Stevens multiplicative factor, and it is known that $120B_4$ equals $10Dq$ [1]. Stevens operators $\hat{O}_4^0, \hat{O}_4^4$ are defined in terms of the orbital angular momentum operators [1, 6] as

$$\hat{O}_4^0 = 35\hat{L}_z^4 - 30\hat{L}^2\hat{L}_z^2 + 25\hat{L}_z^2 - 6\hat{L}^2 + 3\hat{L}^4 \qquad (Eq.S2)$$

$$\hat{O}_4^4 = \frac{1}{2}\left[\hat{L}_+^4 + \hat{L}_-^4\right] \qquad (Eq.S3)$$

Setting $B_4$ as 1, the cubic CEF Hamiltonian for $|L = 2, m_L\rangle$ basis is given by

$$\hat{\mathcal{H}}_{CEF} = \begin{bmatrix} 12 & 0 & 0 & 0 & 60 \\ 0 & -48 & 0 & 0 & 0 \\ 0 & 0 & 72 & 0 & 0 \\ 0 & 0 & 0 & -48 & 0 \\ 60 & 0 & 0 & 0 & 12 \end{bmatrix} \qquad (Eq.S4)$$

where each operator has been normalized by $\hbar$. We would like to note that by setting $|B_4| = 1$, all energy eigenstates will be given in terms of $B_4$ while the positive sign is due to the $d^6$ electron configuration of $Fe^{2+}$, producing a triplet and not a doublet ground state like $Cu^{2+}$ [1]. After diagonalization, the ground state is split into a triplet ground state ($T_{2g}$) and a doublet first excited state ($E_g$), where $\Delta(T_{2g} \to E_g) = 120 B_4$, which is equal to $10Dq$.

$$\begin{bmatrix} -48 & 0 & 0 & 0 & 0 \\ 0 & -48 & 0 & 0 & 0 \\ 0 & 0 & -48 & 0 & 0 \\ 0 & 0 & 0 & 72 & 0 \\ 0 & 0 & 0 & 0 & 72 \end{bmatrix} \qquad (Eq.S5)$$

Utilizing the diagonalized crystal field Hamiltonian above, we can obtain a transformation matrix $\mathcal{U}$ of the form

$$\mathcal{U} = \begin{bmatrix} 0 & 0.71 & 0 & 0.71 & 0 \\ 0 & 0 & 1 & 0 & 0 \\ 0 & 0 & 0 & 0 & 1 \\ -1 & 0 & 0 & 0 & 0 \\ 0 & -0.71 & 0 & 0.71 & 0 \end{bmatrix} \qquad (Eq.S6)$$

where columns of $\mathcal{U}$ are the eigenvectors corresponding to each eigenvalue. The transformation matrix $\mathcal{U}$ rotates operators from the $|L = 2, m_L\rangle$ basis to a $|\phi_{CEF}\rangle$ basis defined by the crystal field eigenvectors as

$$\hat{\mathcal{O}}_{|\phi_{CEF}\rangle} = \mathcal{U}^{-1} \hat{\mathcal{O}}_{|L, m_L\rangle} \mathcal{U} \qquad (Eq.S7)$$

Projecting the $L_z$ operator from the $|L = 2, m_L\rangle$ basis to the $|\phi_{CEF}\rangle$ basis via Eq. S7, one obtains

$$\mathcal{U}^{-1} \hat{L}_z \mathcal{U} = \left[ \begin{array}{ccc|cc} -1 & 0 & 0 & 0 & 0 \\ 0 & 0 & 0 & 2 & 0 \\ 0 & 0 & 1 & 0 & 0 \\ \hline 0 & 2 & 0 & 0 & 0 \\ 0 & 0 & 0 & 0 & 0 \end{array} \right] \qquad (Eq.S8)$$

In this case, the left upper 3×3 block matrix is equal to the $L_z$ operator of $|\tilde{L} = 1, m_{\tilde{L}}\rangle$ basis just with a different sign. So, this orbital triplet ground state ($T_{2g}$) in $|\phi_{CEF}\rangle$ basis can be projected as $\tilde{L} = 1$ state, which is known as T-P equivalence [7,8]. Since these multiplet states are separated by a value of the order of eV, we can reasonably assume the orbital triplet $^5T_{2g}$ state (i.e., $\tilde{L} = 1$ and $S = 2$) as a starting point when considering perturbations with further trigonal distortions and spin-orbit coupling.

Now, we can write down a simplified Hamiltonian for multiplet calculations with a $^5T_{2g}$ state as

$$\mathcal{H} = \mathcal{H}_{SOC} + \mathcal{H}_{tri} = \lambda \tilde{\mathbf{L}} \cdot \mathbf{S} + \Delta_{trig}\left(\tilde{L}_z - \frac{2}{3}\right) \qquad (Eq.S9)$$

where $\lambda$ is the spin-orbit coupling and $\Delta_{trig}$ is the trigonal crystal field. Unlike Quanty multiplet calculation[9], this simplified Hamiltonian only considers the spin-orbit coupling and the trigonal distortion. Using the 15 basis functions of $(2\tilde{L}+1) \times (2S+1)$ in the $|\tilde{L},S\rangle = |\tilde{L}=1\rangle_L \otimes |S=2\rangle_S$ basis, we calculate the ground state of FePS$_3$ by numerically diagonalizing the 15×15 matrix. By setting $\lambda = 13$ meV and $\Delta_{trig} = -10$ meV for the calculation, as in our Quanty multiplet simulation, the Hamiltonian is given by

| $|\tilde{L},S\rangle$ | $\|1,2\rangle$ | $\|1,1\rangle$ | $\|1,0\rangle$ | $\|1,-1\rangle$ | $\|1,-2\rangle$ | $\|0,2\rangle$ | $\|0,1\rangle$ | $\|0,0\rangle$ | $\|0,-1\rangle$ | $\|0,-2\rangle$ | $\|-1,2\rangle$ | $\|-1,1\rangle$ | $\|-1,0\rangle$ | $\|-1,-1\rangle$ | $\|-1,-2\rangle$ |
|---|---|---|---|---|---|---|---|---|---|---|---|---|---|---|---|
| $\langle 1,2\|$ | 16 | 0 | 0 | 0 | 0 | 0 | 0 | 0 | 0 | 0 | 0 | 0 | 0 | 0 | 0 |
| $\langle 1,1\|$ | 0 | 3 | 0 | 0 | 0 | 18.38 | 0 | 0 | 0 | 0 | 0 | 0 | 0 | 0 | 0 |
| $\langle 1,0\|$ | 0 | 0 | -10 | 0 | 0 | 0 | 22.52 | 0 | 0 | 0 | 0 | 0 | 0 | 0 | 0 |
| $\langle 1,-1\|$ | 0 | 0 | 0 | -23 | 0 | 0 | 0 | 22.52 | 0 | 0 | 0 | 0 | 0 | 0 | 0 |
| $\langle 1,-2\|$ | 0 | 0 | 0 | 0 | -36 | 0 | 0 | 0 | 18.38 | 0 | 0 | 0 | 0 | 0 | 0 |
| $\langle 0,2\|$ | 0 | 18.38 | 0 | 0 | 0 | 0 | 0 | 0 | 0 | 0 | 0 | 0 | 0 | 0 | 0 |
| $\langle 0,1\|$ | 0 | 0 | 22.52 | 0 | 0 | 0 | 0 | 0 | 0 | 0 | 18.38 | 0 | 0 | 0 | 0 |
| $\langle 0,0\|$ | 0 | 0 | 0 | 22.52 | 0 | 0 | 0 | 0 | 0 | 0 | 0 | 22.52 | 0 | 0 | 0 |
| $\langle 0,-1\|$ | 0 | 0 | 0 | 0 | 18.38 | 0 | 0 | 0 | 0 | 0 | 0 | 0 | 22.52 | 0 | 0 |
| $\langle 0,-2\|$ | 0 | 0 | 0 | 0 | 0 | 0 | 0 | 0 | 0 | 0 | 0 | 0 | 0 | 18.38 | 0 |
| $\langle -1,2\|$ | 0 | 0 | 0 | 0 | 0 | 0 | 18.38 | 0 | 0 | 0 | -36 | 0 | 0 | 0 | 0 |
| $\langle -1,1\|$ | 0 | 0 | 0 | 0 | 0 | 0 | 0 | 22.52 | 0 | 0 | 0 | -23 | 0 | 0 | 0 |
| $\langle -1,0\|$ | 0 | 0 | 0 | 0 | 0 | 0 | 0 | 0 | 22.52 | 0 | 0 | 0 | -10 | 0 | 0 |
| $\langle -1,-1\|$ | 0 | 0 | 0 | 0 | 0 | 0 | 0 | 0 | 0 | 18.38 | 0 | 0 | 0 | 3 | 0 |
| $\langle -1,-2\|$ | 0 | 0 | 0 | 0 | 0 | 0 | 0 | 0 | 0 | 0 | 0 | 0 | 0 | 0 | 16 |

$(Eq.S10)$

Diagonalizing the simplified Hamiltonian yields the doublet ground state as

$$|GS\rangle_+ = 0.8325\,|-1\rangle_L \otimes |2\rangle_S - 0.4712\,|0\rangle_L \otimes |1\rangle_S + 0.2914\,|1\rangle_L \otimes |0\rangle_S \qquad (Eq.S11)$$

$$|GS\rangle_- = -0.8325\,|1\rangle_L \otimes |-2\rangle_S + 0.4712\,|0\rangle_L \otimes |-1\rangle_S - 0.2914\,|-1\rangle_L \otimes |0\rangle_S \qquad (Eq.S12)$$

From this ground state wavefunction, we obtain an unquenched orbital moment $|\langle L_z\rangle| = 0.6081$, which is close to the value from Quanty. Hence, our simplified CEF multiplet calculation reveals a significant unquenched orbital moment, consistent with the Quanty calculation. Moreover, the ground state wavefunction is not separable into distinct orbital and spin Hilbert space, indicating spin-orbital entanglement. We think that our multiplet model simulation is a simplified yet microscopic physical picture capturing all the essence of the ground state, in particular, the entanglement of the ground state wavefunction.

**Table S1.** Comparison of orbital expectation value with the different model approaches.

| Model | $|\langle L_z\rangle|$ |
|---|---|
| CEF simulation with Eq S3 | 0.6081 |
| Quanty multiplet simulation | 0.7005 |

**Branching ratio and spin-orbit effect in FePS$_3$.**

We tried to extract the spin-orbit coupling (SOC) effect in FePS$_3$ from XA spectra using a branching ratio $BR = I(L_3)/I(L_2)$. The quantity $I(L_{2,3})$ is the integrated intensity of the "white line," (colored area in Fig. S8) an intense absorption spectrum in the Fe $L_{2,3}$ edge. The $BR$ is theoretically related to the expectation value of the spin-orbit operator (or spin-orbit expectation value) $\langle \boldsymbol{L} \cdot \boldsymbol{S} \rangle$ as follow [7, 8].

$$BR = \frac{I(L_3)}{I(L_2)} = \frac{2+r}{1-r} \qquad (Eq.\,S13)$$

where $r = \langle \boldsymbol{L} \cdot \boldsymbol{S} \rangle / \langle n_h \rangle$ and $\langle n_h \rangle$ is the average number of 3d holes. Note that $BR$ is usually equal to 2 when the spin-orbit effect is absent [10-13]. This close connection between $BR$ and spin-orbit expectation value has been demonstrated as a strong indicator of significant spin-orbit entanglement in 4d and 5d compounds [10-14].

The $BR$ value obtained from the experimental XA spectra at 65 K is 3.84 and $\langle \boldsymbol{L} \cdot \boldsymbol{S} \rangle = 1.38$. To check the impact of SOC, we simulate the $BR$ with several conditions using the Quanty multiplet calculation (see Table S2). First, when we only consider trigonal distortion, $BR = 2.60$, slightly deviated from the 2, and $\langle \boldsymbol{L} \cdot \boldsymbol{S} \rangle = 0.67$. The non-zero spin-orbit expectation value $\langle \boldsymbol{L} \cdot \boldsymbol{S} \rangle$ comes from the unquenched orbital moment. Next, we consider only SOC, which results in the $BR = 3.5$ and $\langle \boldsymbol{L} \cdot \boldsymbol{S} \rangle = 1.33$. In this case, the spin-orbit expectation value solely comes from the spin-orbit entangled state. Finally, we obtain $BR = 3.33$ and $\langle \boldsymbol{L} \cdot \boldsymbol{S} \rangle = 1.23$ when considering both trigonal distortion and SOC, which implies that the ground state of $FePS_3$ is close to the spin-orbit entangled limit.

We would like to stress that $BR$ is proportional to the $\langle \boldsymbol{L} \cdot \boldsymbol{S} \rangle$, not the SOC constant $\lambda$. It is also known that strong spin-orbit interaction does not always guarantee a large $BR$ [11]. So, a large $BR$ far from 2 shows that $FePS_3$ has a strong spin-orbit entanglement, as we pointed out in the ground state wavefunction calculation.

**Table S2.** Comparison of the $BR$ and the spin-orbit expectation value $\langle \boldsymbol{L} \cdot \boldsymbol{S} \rangle$ of experiment and Quanty multiplet simulation. For the trigonal distortion and spin-orbit coupling, we consider $D_{t2g} = -10$ meV and $\lambda = 13$ meV.

|  | BR | $\langle n_h \rangle$ | $\langle \boldsymbol{L} \cdot \boldsymbol{S} \rangle$ (units in $\hbar^2$) |
|---|---|---|---|
| **Experiment** | 3.84 | 3.62* | 1.38 |
| Only $D_{t2g}$ | 2.60 | 4 | 0.67 |
| $D_{t2g}$ & SOC | 3.33 | 4 | 1.23 |
| Only SOC | 3.50 | 4 | 1.33 |

*The hole value of experimental data is adapted from the Quanty calculation's result

## Charge transfer from ITO substrate to 1st layer of FePS3

We use a commercial indium tin oxide (ITO)/Si substrate to avoid charging effects because the $FePS_3$ is strongly insulating; we also tried other substrates, such as Si oxide/Si, with much more severe problems due to charging effects. However, even in the case of ITO, the work function mismatch between the p-type semiconductor $FePS_3$ [15,16] and the metallic ITO [17] results in a Schottky barrier, inducing electron transfers from ITO to $FePS_3$. This interfacial charge transfer effect renders the XA spectrum of the monolayer significantly different from that of the 20 ML (Fig. S5 a). The A, B, and C pre-peaks are visibly suppressed while a new post-peak appears. The XMLD effect is reduced and remains within the noise levels of the spectra (Fig. S5 b).

To verify the charge transfer effect from the metallic substrate, we change the parameters related to the ligand-metal charge transfer $H_{lmct}$ and the ligand's crystal field $H_L$ from the Quanty multiplet calculation in the Methods. By reducing the charge transfer energy from ligands to the *3d* metal ion, we could imitate the charge transfer effect from the substrate, in agreement with our experimental results. The modified input parameters are summarized in Table S3.

**Table S3**. Summary of parameters

| Hamiltonian | Parameters |
| --- | --- |
| *Atomic terms $H_d$ | $U_{dd}$ = 2.0 eV $U_{pd}$ = 2.5 eV |
| **Crystal field $H_c$ | 10Dq = 0.4 eV $D_{t2g}$ = -0.01 eV |
| ***Ligand-metal charge transfer $H_{lmct}$ | $\Delta$ = -6.0 eV $V_{a1g}$ = 1.5 eV $V_{eg\pi}$ = 1.5 eV $V_{eg\sigma}$ = 2.0 eV |
| Ligand crystal field $H_L$ | 10Dq$^{lig}$ = 0 eV $D_{t2g}^{lig}$ = 0 eV |

*Slater integrals are reduced to 80% of the Hartree-Fock value with the spin-orbit coupling constant of $\xi$ = 0.05 eV.

**$D_{t2g}$ is defined as E ($e_g^\pi$) - E ($a_{1g}$), hence the negative $D_{t2g}$ term reflects the trigonal elongation.

***These terms are for the charge transfer effect of sulfur ligand-*3d* iron metal.

For monolayer, the value of *3d* electrons occupation number is determined to be $<N_{3d}>$ = 6.77 while the bulk value was found to be $<N_{3d}>$ = 6.38. The additional electron found in the monolayer sample comes from the substrate due to the charging effect. It is expected to occupy the low-lying $e_g^\pi$ state, reducing the magnetic anisotropy as found in our analysis for the monolayer sample.

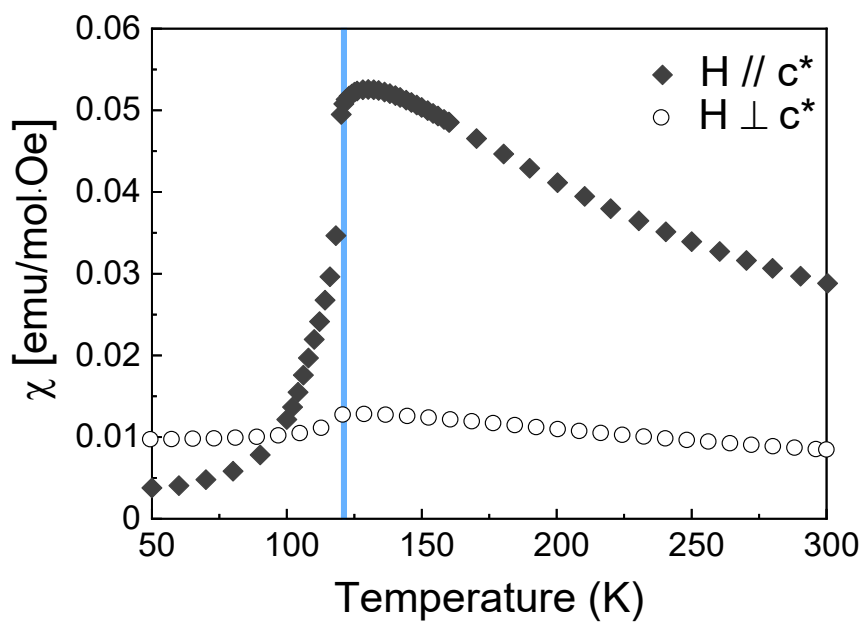

**Fig S1**. Magnetic susceptibility of bulk FePS$_3$.

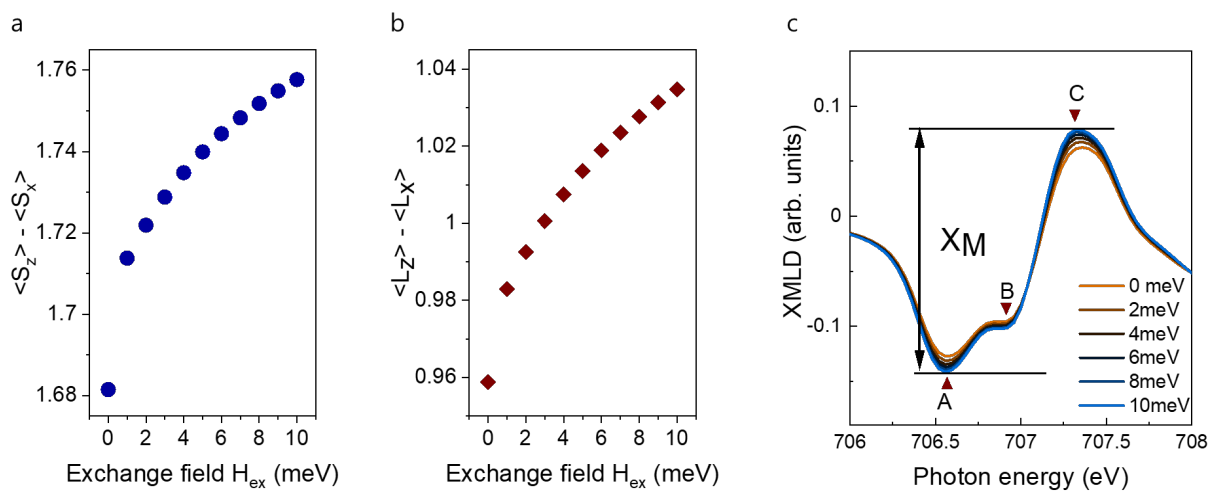

**Fig S2**. Exchange field effect on magnetic anisotropy and spectra in the multiplet calculation. Simulated **a** spin and **b** orbital anisotropy as a function of exchange field H$_{ex}$ from 0 meV to 10 meV. **c** XMLD spectra with varying H$_{ex}$.

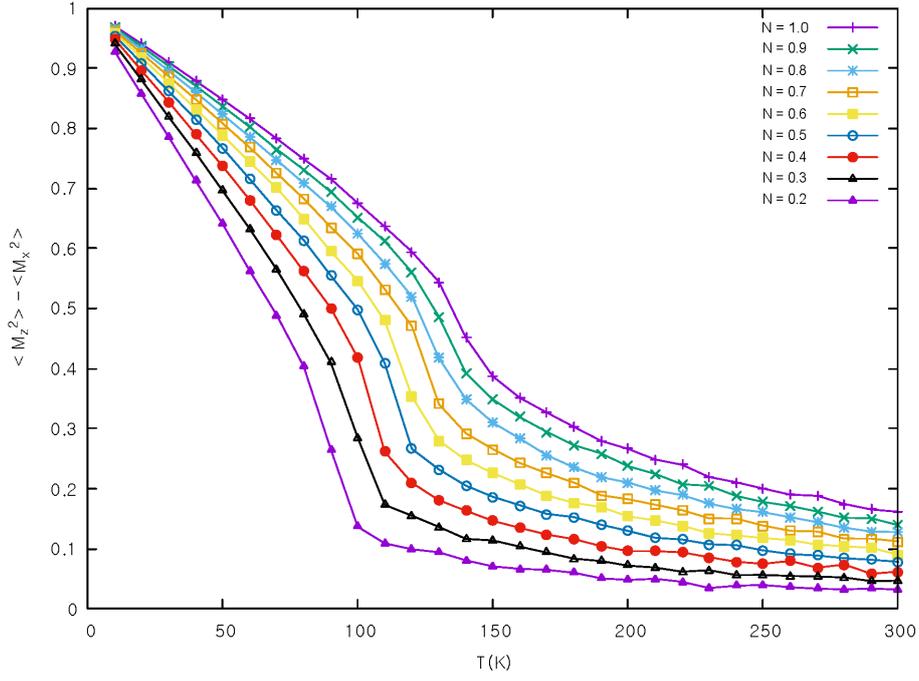

**Fig S3**. Magnetic anisotropy as a function of temperature simulated by the Monte-Carlo method. The legend shows scaling factor N where $K_N = K \times N$ with K = 22 meV/Fe ion.

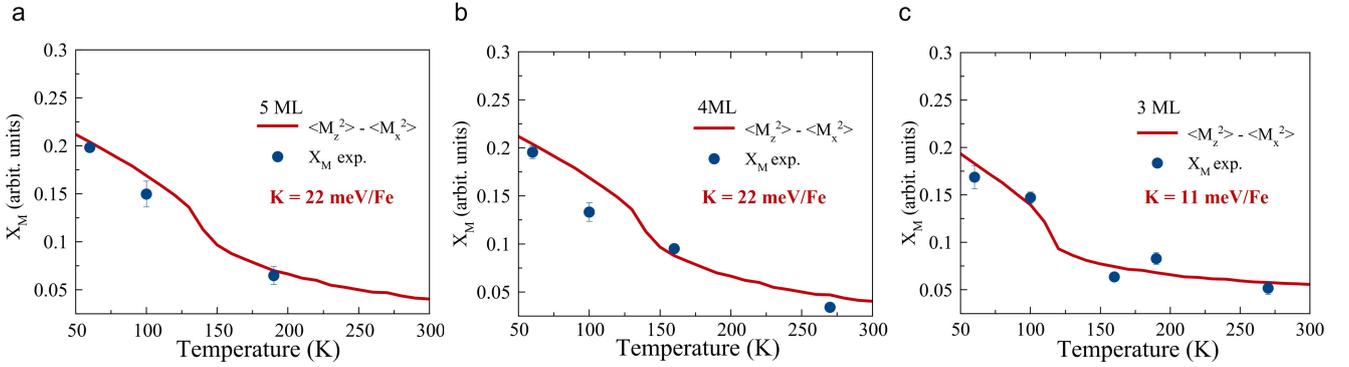

**Fig S4**. Magnetic anisotropy as a function of thickness. Fitting of simulated magnetic anisotropy from the Monte-Carlo simulation with the experimental data $X_M$ in **a** 5 ML, **b** 4ML, and **c** 3ML.

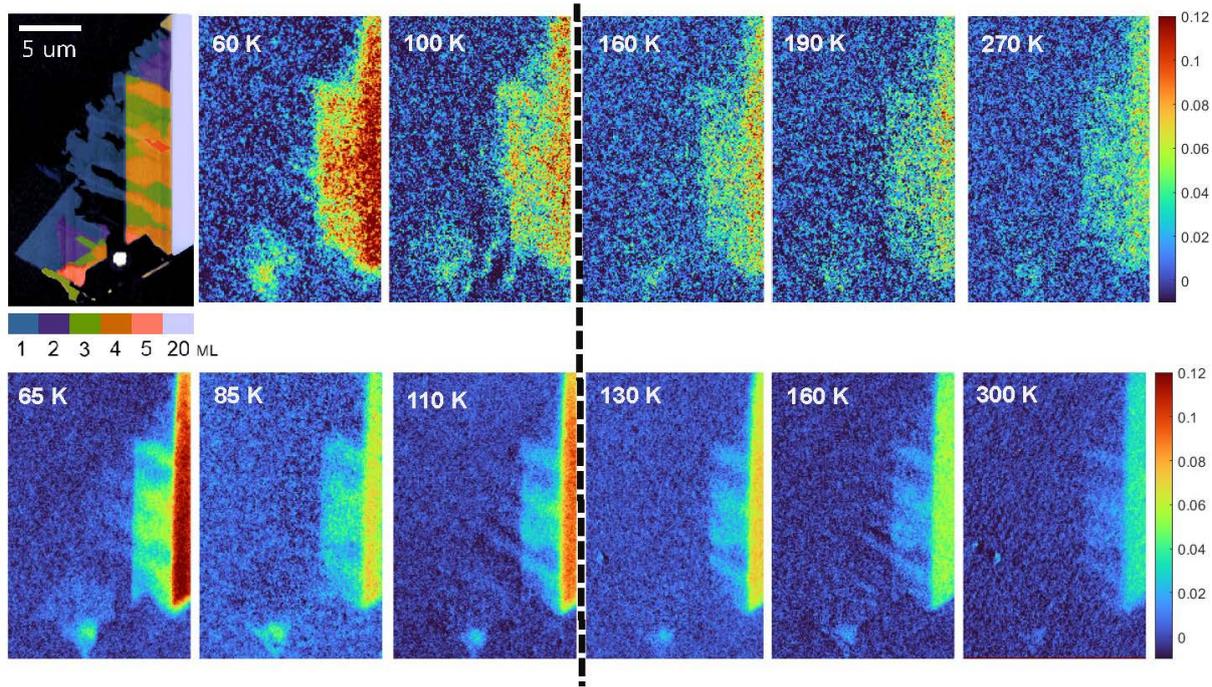

**Fig S5.** Temperature-dependent XMLD asymmetry map. The first far-left image in upper half is the false-colored AFM image for thickness information while the other figure in the upper half is for the first scan. And all the figures in the bottom half is for the second scan with increasing temperature. The black dotted line indicates the $T_N$ = 120 K.

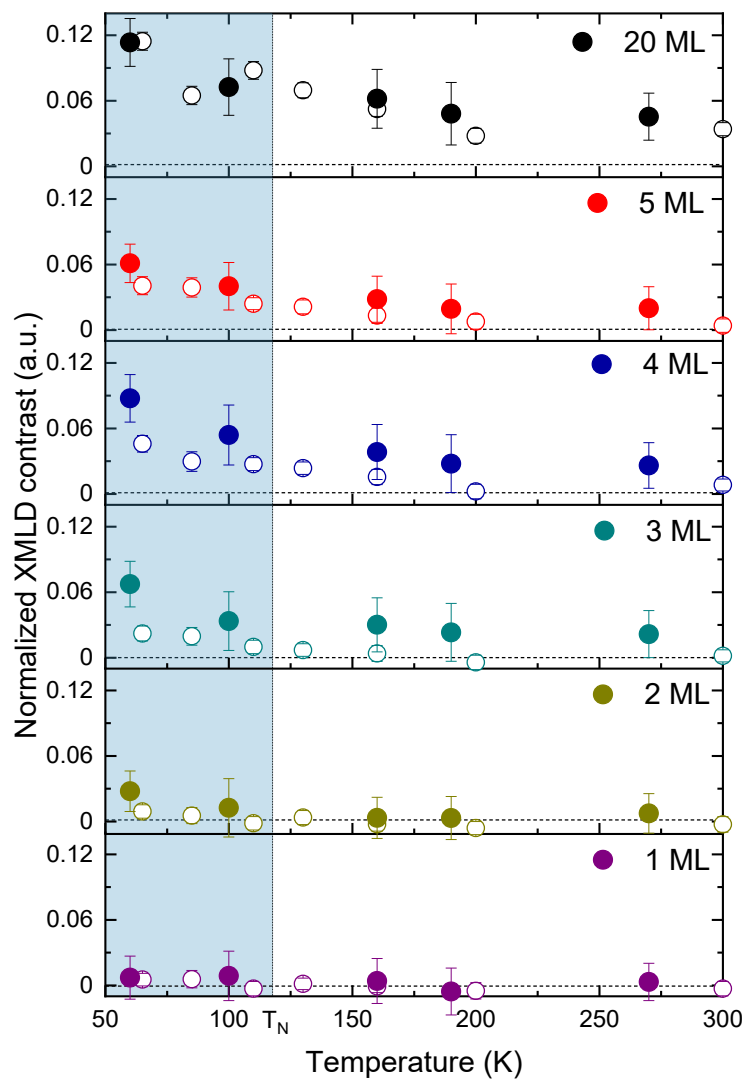

**Fig S6.** XMLD contrast value from the XMLD asymmetry map. The filled circles are value from the first scan while the vacant circles are from the second scan's contrast amplitude.

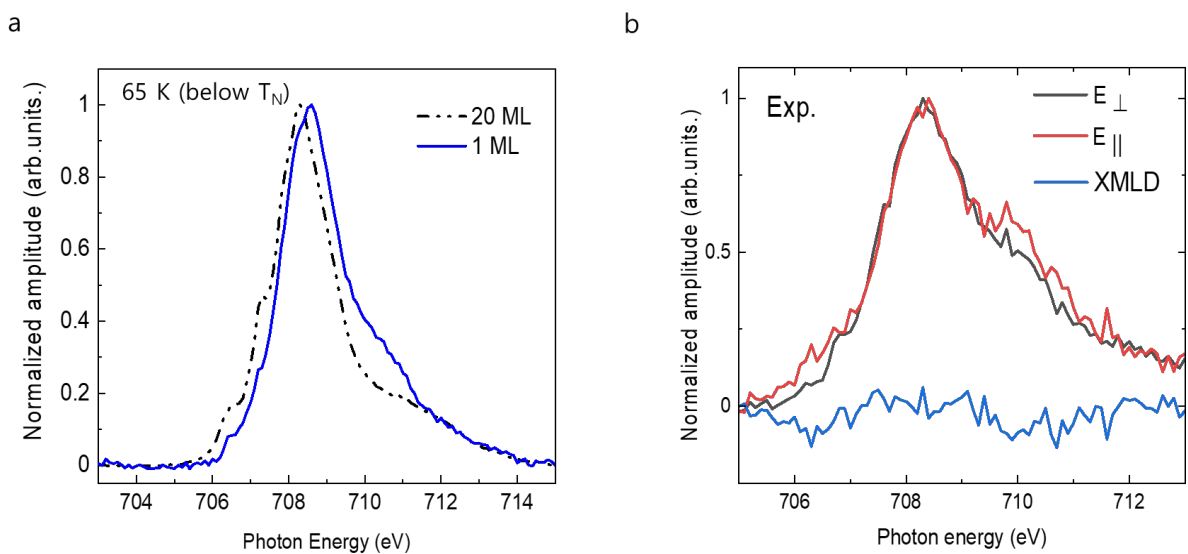

**Fig S7**. Charge transfer from ITO substrate to 1st layer of FePS$_3$. **a** The deviation of the XA spectrum of monolayer from that of 20 ML. **b** Experimental XA spectra with linear polarization and linear dichroic spectra of the monolayer.

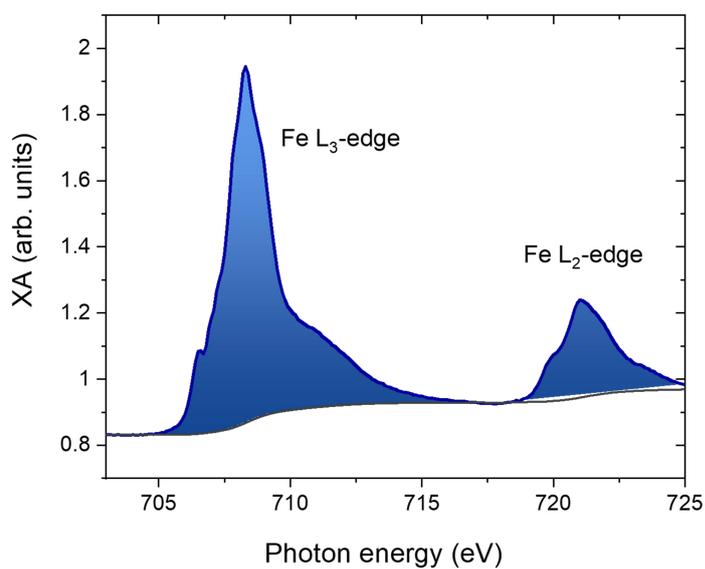

**Fig S8**. Branching ratio of FePS$_3$. Blue-shadowed areas are "white line" of experimental isotropic XA spectrum.

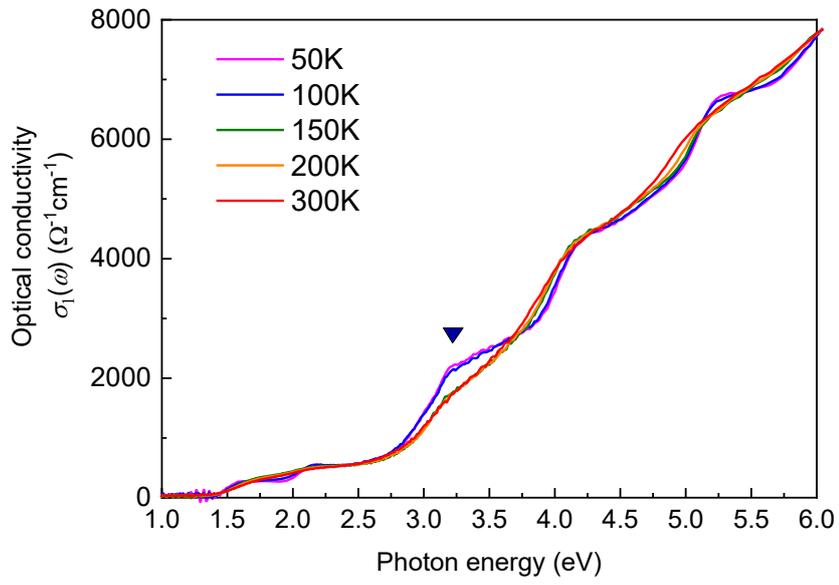

**Fig S9**. Temperature-dependent optical conductivity measurements.